\documentstyle[twocolumn,amssymb,aps,graphicx]{revtex}
\begin{document}
\title{  Correlated Persistent Tunneling Currents in  Glasses}
\author{{\large
 Stefan Kettemann (1,2 *), Peter Fulde (1), Peter Strehlow (3) }
}
\address{ (1) Max-Planck Institut
 f\" ur Physik Komplexer Systeme, 
 N\" othnitzerstr. 38, 01187 Dresden,\\ 
 (2)I. Institut f. Theoretische Physik, 
 Universit\" at Hamburg, Jungiusstr. 9, 20355 Hamburg,\\
 (3)  Physikalisch-Technische Bundesanstalt Berlin, 
 Abbestra{\ss e} 2-12,  10587 Berlin }
\date{\today }
\maketitle
        
\begin{abstract}
Low temperature properties of glasses are derived within a generalized
tunneling model,
considering the motion of charged particles on a closed path in a 
double-well potential. 
The presence of a magnetic induction field $\bbox{B}$
violates the time reversal invariance due to the 
Aharonov-Bohm phase, 
and leads to flux periodic energy levels. 
At low temperature, this effect is 
shown to be strongly enhanced
by  dipole-dipole and elastic interactions 
between  tunneling systems 
and becomes measurable.
Thus, the recently observed strong sensitivity 
of the electric permittivity to weak magnetic fields can be explained. 
In addition, superimposed oscillations 
as function of the magnetic field are predicted.  
PACS- numbers: 61.43 Fs, 66.35.+a,  77.22.Ch,  03.65.Bz
\end{abstract}

At low temperature, glasses exhibit a variety of surprising 
properties of considerable theoretical interest. 
They are attributed to the existence of low-energy 
excitations present in nearly all 
amorphous solids and disordered crystals \cite{Phillips}. In 
the standard tunneling model (STM)
these excitations are described on a phenomenological 
basis by noninteracting two-level tunneling systems (TLS's)\cite{Anderson}. 
Such a TLS  
can be approximately treated as a particle moving 
in a double-well potential. At sufficiently low temperature 
only the ground states in the two wells are relevant, and the system
is effectively restricted to the two-dimensional Hilbert space. 
Thus, the Hamiltonian 
of a TLS has the form $ H_0 = (1/2) ( \Delta \sigma_z - \Delta_0 
\sigma_x ) $, where $\Delta$ and $\Delta_0$ are  the
asymmetry energy and the tunneling matrix element, respectively. 
Because of the random structure of glasses it is assumed that 
$ \Delta$ and $\Delta_0$ exhibit a wide distribution according to
$P ( \Delta, \Delta_0 ) = \bar{P} /\Delta_0$, where $\bar{P}$ is a constant.
Treating the coupling of TLS's with external acoustic and electric fields 
as a weak perturbation of $H_0$, the STM has successfully explained many of
the anomalous thermal, acoustic and dielectric 
low-temperature properties of glasses.
Deviations from the predictions of the STM \cite {review} have been usually reduced to
interaction between TLS's \cite{Yu}. 
Based on the 
spin-Boson-Hamiltonian, the elastic interaction between TLS's
can be investigated in a nonperturbative manner \cite{Kassner}. 
In that way it can be demonstrated that the strong TLS-phonon coupling
essentially leads to a renormalization 
of the tunneling parameters, $\bar{P}$ 
and $\Delta_{ \rm 0 min}$, and thus to the quasiparticle picture of 
phonon-dressed TLS's.          
It must be asked, however, whether interaction can also lead to an
observable action of electromagnetic fields on the quantum-mechanical
state of TLS's. This is motivated by the recently reported strong 
sensitivity of the electric permittivity (dielectric constant) 
to weak magnetic fields in multicomponent glasses at ultra-low 
temperature \cite{Strehlow}. 
The purpose of the present paper is therefore, to investigate the 
influence of static electromagnetic fields on the 
energy spectrum of a TLS, and how 
it is modified by the coupling between them. 
 
In order to estimate electromagnetic field effects upon the electric
permittivity of glasses in a classical approach, we can start from 
general thermodynamic principles, treating glasses as 
isotropic non-viscous dielectrics devoid of any free charges and free currents. The 
constitutive quantities of such a material can depend on 
mass density $\rho$, velocity $\bbox{\dot x}$, temperature $T$, 
temperature gradient ${\bf \nabla } T$, 
electric field $\bbox{E}$ and magnetic flux density $\bbox{B}$. 
The general form of the constitutive equation for a constitutive quantity 
such as the polarization $\bbox{P} = \bbox{P}[
\rho,\bbox{\dot x},T,\nabla T,\bbox{E},\bbox{B}]$
is restricted by the principle of material frame indifference (or objectivity, PMO) 
and the entropy principle \cite{Thermo}.
With respect to Euclidean transformations,
$\bbox {x}^{*} = \bbox {O(t)}\bbox{x} + \bbox{b(t)}$, where $\bbox{b}(t)$ and $\bbox{O}(t)$ 
determine the translation of two frames and the relative rotation of their axes, respectively,
the PMO states that the constitutive equations must be invariant under change
of frame. This means that $\bbox{\dot x}$ and $\bbox{E}$ cannot occur separately
as variables, but only in combination 
as electromotive intensity $\bbox{\cal E} = 
\bbox{E} + \bbox{\dot x}\times\bbox{B}$.
For the polarization the functional equation
$\bbox{P}[\rho,T,\bbox{O}\nabla T,\bbox{O}\bbox{\cal E},(\rm{det}\bbox{O})\bbox{O}\bbox{B}] = 
\bbox{O}\bbox{P}[\rho,T,\nabla T,\bbox{\cal E},\bbox{B}]$
must be satisfied for all orthogonal matrices $\bbox{O}$. 
Recall that $\bbox{P}$ and $\bbox{\cal E}$ are polar vectors,
while $\bbox{B}$ is an axial vector. The general solution of the functional equation leads to
the representation for the polarization, whose equilibrium part ($\nabla T = 0$)
is given by $\bbox{P}/\epsilon_0 = \chi_1\bbox{\cal E} + \chi_2(\bbox{B}\cdot\bbox{\cal E})
\bbox{B} + \chi_3\bbox{B}\times\bbox{\cal E}$.
The susceptibilities $\chi_{\alpha}$
can still depend on the scalars
$\rho,T,{\cal E}^2,B^2$ and $(\bbox{\cal E}\cdot\bbox{B})^2$.
As a further restriction,
it follows from the entropy principle that $\chi_3 = 0.$
Consequently, the electric permittivity $\epsilon=1+\chi_1+\chi_2B_z^2$ of a rigid 
($\bbox{\dot x}=0$) dielectric and magnetizable glass, determined by capacitance
measurement in a plate condenser with $\bbox{E}=(0,0,E_z)$, 
is expected to depend on $E_z^2$ and $B^2$ (and besides on
$B^4$, if $\bbox{ B}||\bbox{E}$). The quadratic field dependence of $\epsilon$ was
indeed reported to be valid for some glasses \cite{naughton}. In general, however, glasses
are properly assumed to be linear dielectrics, so that $\epsilon(\rho,T)$ does not
depend on the electromagnetic field. Several glasses do not show any measurable changes
of their electric permittivity up to high magnetic fields and temperatures down to a few mK 
\cite{frossati}. The recently observed magnetic field dependence of $\epsilon$ 
in multicomponent glasses at ultra-low temperature \cite{Strehlow}, however,
is completely different in nature from the magnetoeffect in nonlinear dielectrics, 
and cannot be derived from thermodynamics assuming glasses 
as simple magnetizable dielectrics. The knowledge of the local field strengths
$\bbox B$ and $\bbox E$ only is not sufficient for the consistent description
of electromagnetic field effects on the quantum-mechanical state of charged
particles \cite{gorkov}.
For that reason the influence of a magnetic field
on the energy spectrum of a TLS should be investigated 
from a microscopic point of view. In order to describe a TLS in a magnetic field the 
three-dimensional motion of a charged particle 
in an electrostatic double-well potential has to be considered.
Imagine a hat like potential with two potential barriers in
azimuthal direction along the rim of the hat as shown in Fig. 1.

\begin{figure}[bhp]\label{fig1}
\includegraphics[width=0.44\textwidth]{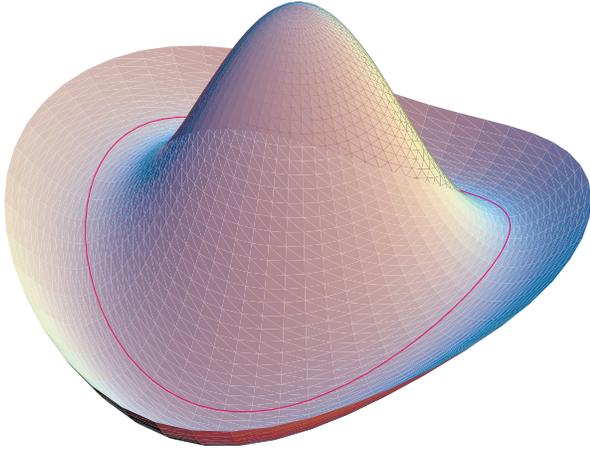}
\caption{Hat potential.
The double-well potential for a charged particle confined 
to a circular path is indicated by the line.
The induction field $\bbox{B}$ can have an arbitrary 
angle with the plane of motion}
\end{figure}  
The Hamiltonian for a non-relativistic and spinless charged particle of mass $m$ 
and charge $q$, which moves in an uniform induction field $\bbox{B}$ and 
in an electrostatic potential $V$, is given by     
\begin{equation}
H_0 = (\bbox{p}-Q \bbox{A})^2/(2m) + V .
\end{equation}
The symmetric gauge for the vector potential $\bbox{A}=(0,B r/2,0)$ 
is appropriate to cylindrical co-ordinates $(r,\theta,z)$.
An uniform induction field $B$ along the $z$ axis forces the charged
particle to circular motion in the perpendicular plane at radii 
$r_n$ with $B \pi  r_n^2 = \phi_0 n$, 
where $\phi_0 = h/Q$ is the flux quantum, and $n$ an arbitrary integer.
The magnetic length $l_B=\sqrt{\hbar/(Q B)}$ characterizes 
the strength of this magnetic confinement and has to be compared with
the radius $r_V$ at which the particle is restrained by   
an electrostatic potential $V(r, \theta, z)$ of the form shown in Fig. 1. 
For weak magnetic fields, $r_V \ll l_B$, the confinement force due to $V$
exceeds the Lorentz force, and Landau quantization does not appear. 
In that case, the problem is reduced to that of an one-dimensional 
ring of radius $r$, with the Hamiltonian
\begin{equation}\label{single}
H_0 =\frac{1}{2 m}(-i \hbar \frac{1}{r}\frac{\partial}{\partial 
\theta} - Q { A_{\theta}} )^2 + V( \theta ) ,
\end{equation}
where $A_{\theta}=B r/2=\phi/( 2 \pi r)$. In the more general case that the
induction field $B$ has an arbitrary angle $\alpha$ with the plane
of motion, the magnetic flux through the circular orbit of a TLS is
given by $\phi=B \pi r^2$cos$\alpha$.

For a double-well potential $V(\theta) = V \cos( 2 \theta )$, 
the Schr\" odinger equation can be reduced by means of the 
gauge transformation 
$\tilde{\varphi}(\theta) =\exp (-i \theta \phi/\phi_0)
\varphi(\theta)$ to the Mathieu equation 
$\partial^2 \tilde{\varphi} / \partial \theta^2
 + ( E/a - (V/a) \cos ( 2 \theta )) \hat{\varphi } =0 $, 
where the energy $E$ and the potential amplitude $V$ 
enter in units of $ a = \hbar^2/(2 m r^2)$.  
Thereby, the magnetic flux is removed from the Hamiltonian, 
and the periodic boundary conditions
$ \varphi ( \theta + 2 \pi ) = \varphi ( \theta )$ 
are substituted by the magnetic flux twisted boundary conditions
$\tilde{\varphi} ( \theta  + 2 \pi  ) =
\exp (- i 2 \pi \phi/\phi_0 ) \tilde{\varphi} ( \theta  ) $. 
Solutions with these boundary conditions do exist 
for particular values of the ratio $E/a$, which yield  the 
energy eigenvalues. They are periodic in the magnetic flux $\phi/\phi_0$ 
as seen in Fig. 2, where the energies $E_-$ and $E_+$ of the
ground and the first excited state are plotted. 

\begin{figure}[bhp]\label{fig2}
\includegraphics[width=0.44\textwidth]{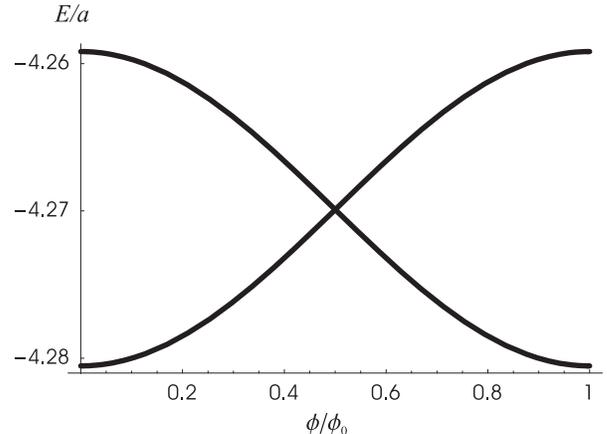}
\caption{The two lowest energy eigenvalues as function of the 
 flux ratio $\phi/\phi_0$ through the circular path 
 of a charged particle in a symmetric double-well potential }
\end{figure}  

For non-zero magnetic flux $\phi$ the energy eigenfunctions composed of the odd and
even Mathieu functions become complex. Thus, due to the magnetic induction field 
both the parity and the time reversal invariance of the TLS are broken.
Each energy eigenstate does
carry a persistent current \cite{hund} of opposite direction, given by
\begin{equation}
I_{\pm} ( \varphi ) =  \partial E_{\pm} ( \phi)/ \partial
\phi.  
\end{equation} 
A net persistent tunneling current results in a magnetic moment of the TLS.     

For more general double-well potentials, 
allowing also an asymmetry $\Delta$ in the energy of the two minima,  
the harmonic approximation of the potential 
can be done around each minimum, in analogy to the STM. 
Then, the ground state and the first excited state of the TLS
can be well approximated by a superposition of 
the ground states of each harmonic oscillator,
if $V \gg \hbar \Omega >\Delta$, 
where $\Omega$ is the oscillator frequency 
and $V \approx (1/8) m \Omega^2 \pi^2 r^2$ is the potential barrier.
The energy eigenvalues are found to be 
$ E_{\pm} =(\hbar \Omega +
\Delta \pm  E_{\rm g}( \phi))/2$, giving the excitation energy (gap),
\begin{equation}
E_{\rm g}(\phi) = 
\sqrt{ \Delta^2 + 
t ( \phi )^2 },
\end{equation}
where $ t ( \phi ) =  \Delta_0 \cos ( \pi
\phi/\phi_0 ),$ with $ \Delta_0\approx\hbar\Omega\exp (  - \lambda ) $
and $\lambda =  2V/(\hbar\Omega) > 1$. 
This is identical to  the STM, when there the tunneling parameter 
$\Delta_0$ is substituted by the magnetic flux dependent 
tunneling splitting $t(\phi)$.
 
The low-temperature thermodynamic properties calculated in the
generalized model of independent TLS's,
which are determined by the energy spectrum $E_{\rm g}(\phi)$, are consequently
periodic functions of the magnetic flux with a period of $\phi_0$.  
Thus, the TLS energy density $e(\phi,T)$ can be obtained by averaging
the excitation energy over the parameters $\Delta$ and $\Delta_0$,
according to
\begin{equation}
e(\phi,T)=\bar{P}\int\frac{\mbox{d} \Delta_0}{\Delta_0}\int \mbox{d} 
\Delta \frac{E_g}
{\exp[E_g/(k_B T)]+1}.
\end{equation}
For temperatures $T \gg \Delta_{\rm 0min}/k_{\rm B}$ the specific heat,
$c(\phi,T)=1/\rho\cdot\partial e(\phi,T)/\partial T$, 
is approximately given by
\begin{equation}
c(\phi,T)=\frac{\pi^2}{6\rho}\cdot\bar{P}k_{\rm B}^2T
\Bigl[1+\ln\Bigl(\frac{2k_{\rm B}T}{\mid t_{\rm min}(\phi)\mid}\Bigr)\Bigr] ,
\end{equation}
where $t_{\rm min}(\phi)=\Delta_{\rm 0min}\cdot\cos(\pi\phi/\phi _0)$ is
the flux dependent minimal tunneling splitting.
The presence of an external electrical field $\bbox E$ 
produces an interaction energy of a TLS with the dipole
moment $\bbox p$ of amount $H_{\rm int}=-\bbox p \cdot\bbox E$. 
Due to $\bbox E$ the asymmetry of the TLS is changed, 
while the change in the tunneling splitting can be neglected for 
$\bf p \cdot\bf E \ll V$. The thermodynamic 
polarization, $\bbox P=-\langle \partial\hat H/\partial\bbox E \rangle$, 
can be derived from the effective Hamiltonian $\hat H=H_0+H_{\rm int}$.
The ensemble average over the different TLS's 
is done by averaging over possible dipole orientations and the parameters
$\Delta$
 and $\Delta_0 $ 
with the measure $\bar{P}\cdot d \Delta d \Delta_0 /\Delta_0$. 
As a result, the resonant part of the electric permittivity 
$\epsilon_{\rm res}$  
becomes in linear response
 ($\bbox P=\epsilon_0(\epsilon-1)\bbox E$),
\begin{equation}
\epsilon_{\rm res}-1=\frac{2\bar{P}p^2}{3\epsilon_0} 
\int_{\mid t_{\rm min}(\phi)\mid}^{E_{\rm max}} \mbox{d} E_{\rm g} 
\frac{\sqrt{E_{\rm g}^2 - t_{\rm min}(\phi )^2}}
{E_{\rm g}^2}\tanh(\frac{E_{\rm g}}{2k_{\rm B}T}).
\end{equation}
The electric permittivity depends on temperature as well as   
 magnetic flux through the minimal tunneling splitting. 
There are maxima in $\epsilon_{\rm res}(\phi,T)$ 
at
 $\phi/\phi_0=(2n+1)/2$,
$n=0,1,2,...,$
 where the lower cutoff of the excitation energy vanishes. 
These maxima become more pronounced as the temperature is lowered.
Thus, the resonant part  $\epsilon_{\rm res}$
 does depend on the magnetic field
at low temperature where the relaxational contribution to the 
permittivity
 $\epsilon_{\rm rel}$ is negligible.
Indeed, the theoretically derived dependence of the electric permittivity, Eq. (7), 
on magnetic field and temperature explains qualitatively the main feature
of the experimental results obtained in multicomponent glasses \cite{hunklinger}
as shown in Fig. 3.
\\
\\
\begin{figure}[bhp]\label{fig3}
\begin{center}
\includegraphics[width=0.47\textwidth]{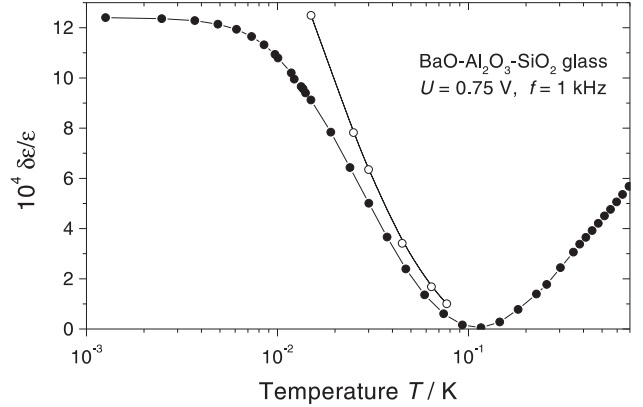}
\caption{Temperature variation of the electric permittivity
$\delta\epsilon/\epsilon = [\epsilon(T)-
\epsilon(T_{\rm 0})]/\epsilon(T_{\rm 0})$ 
of a BaO-Al
$_{2}$O$_{3}$-SiO$_{2}$
glass measured at a frequency of 1 kHz and a voltage of
0.75 V (data from Ref. (12)). $T_{\rm 0}$=113 mK
 was taken as 
a reference. The solid circles represent the data in zero field, 
whereas the open circles represent the maximal measured increase of 
$\epsilon$ 
due to a magnetic field. As shown by the solid lines, 
the data can be described by Eq.(7) in the resonant regime below 100 mK 
with $t_{\rm min}(\phi=0)/{\rm k_B}=12.2$ mK
 (lower line) 
and $t_{\rm min}(\phi=\phi_{0}/2)=0$ 
(upper line), respectively.
Above 100 mK the relaxational part $\epsilon_{\rm rel}$
  in $\epsilon$
is additionally taken into account.}
\end{center}
\end{figure}  

In the low-temperature resonant regime, the measured temperature dependence
of $\epsilon(T,\phi=0)$ 
can be well described by Eq.(7) assuming 
$\bar P p^2/\epsilon_0=1.03\cdot 10^{-2}$ 
and $\Delta_{\rm 0min}/k_{\rm B}=12.2$ mK. 
The deviation from the logarithmic temperature dependence of 
$\epsilon_{\rm res}$
(dielectric saturation \cite{rogge}) is 
lifted   at magnetic fields where 
$\epsilon_{\rm res} (T,\phi=\phi_{0}/2)$
 becomes maximal. In that case, the lower cutoff $t_{\rm min}(\phi)$
 vanishes, and $\epsilon$
 varies logarithmically with temperature.
The remarkable result of the experiments consists in the fact that both
the calculated logarithmic temperature dependence of 
$\epsilon_{\rm res}$ and its
slope relative to the relaxational part $\epsilon_{\rm rel}$ 
of (-2):1 at
higher temperature is achieved even at weak magnetic fields of about 0.1 T.
The experimentally observed maximum in $\epsilon(T,\phi(B))$ 
at $B \approx 0.1$ T, 
which slightly depends on temperature,
 requires to assume for $r \approx 2\cdot 10^{-10}$ m 
the TLS charge of $Q \approx 4\cdot 10^5|\rm{e}|$,
 where e is the elementary charge.  
Before we discuss the origin of such a large value of $Q$ 
resulting apparently
from interactions of TLS's, we want to mention that the 
experiments described in 
\cite{hunklinger}, which were stimulated by our theoretical findings, 
have confirmed
a corresponding flux periodic behaviour of both the specific heat and the electric
permittivity. 
It should be noted that $\phi$ in Eqs. (6) and (7) has to be
interpreted as an effective magnetic flux, and an averaging 
over orientations and 
charges of TLS's must be carried out in order to analyze
 precisely the oscillatory
behaviour of $c(T,\phi)$ 
and $\epsilon(T,\phi)$ in magnetic fields.

In order to explain the large values of $Q$, we have to consider  
the excitation spectrum of coupled rings.
 The Hamiltonian 
is given by
\begin{eqnarray}\label{multi}
 H &=& \sum_i H_{0 i} + \sum_{i \neq j} 
g( \theta_i, \theta_j)/r_{i j}^3, 
\end{eqnarray}   
 The first term is the sum of the  Hamiltonians of the uncoupled rings,                      Eq. (\ref{single}).
 The second term is the (dipole-dipole or elastic) interaction energy 
decaying with the distance $r_{ij}$ between two TLS's $i$, $j$,
 and depends
 on the orientations 
of the dipole moments 
 as parametrized by the angles $\theta_i$.
We find  \cite{kett} when the interaction energy exceeds
 the typical kinetic energy scale in each ring,
that the excitation spectrum of  strongly 
coupled rings  equals        
that of   one      ring with an effective charge $Q$,
which is   the sum of the charges of the rings.
 This is     intuitively clear, 
since then   the motion of the tunneling 
 particles
 is governed by the interaction between them, 
making all degrees of freedom massive apart from 
  their center of mass motion.
 Thus, we have identified
a possible origin for large values of Q. 
 Similar situations, where the tiny magetic response of  microscopic 
 entities like a molecule is enhanced  to a macroscopic 
 magnetic field effect 
 due to correlations between them,  
 are known. One example is 
 the Frederiks transition in nematic liquid crystals\cite{degennes}.
 Assuming for the tunneling
parameters the feasible values 
 $\bar P=10^{45} \rm{J}^{-1}\rm{m}^{-3}$,
 $E_{\rm{max}}/\rm k_B=5$ K,
 $\Delta_{\rm{0min}}/\rm{k_B}=10^{-6}$
 K, 
 the average distance
between TLS's can be estimated to be  $r_{nn} = 10^{-8}$ 
m.
 Thus, a   number $10^5$ of coupled rings 
 implies mesoscopic coherence lengths,
 on the order of 
$\mu$m.
 For an averaged
dipole moment
 $p=2|e|\cdot10^{-10}$ m, 
the dipole-dipole interaction energy is
then $g/{\rm{k_B}}/r_{nn}^3 \approx 100$ mK.
 This is exactly the temperature range
 in which the 
magnetic flux effects become observable. 
The obtained energy  
spectra for strongly coupled rings suggest the introduction of quasiparticles 
whose orbits are pierced by a flux with flux periodicity $\phi_0=h/Q   $. 
They are  excitations 
of strongly coupled TLS's 
 with renormalized tunneling parameters. 
Due to interactions the lowest quasiparticle excitation energy 
 can be considerably changed. For example, for two coupled rings with
$U > t(\phi)$ 
the splitting is reduced to  $t(\phi)^2/U$. 
     
However, there exists another phenomenon which contributes to changes in
the energy spectrum and deserves special attention. It can be visualized by 
investigating in detail the energy levels of two interacting
 TLS's\cite{kett}. The dipole
moment of an asymmetrical TLS increases with decreasing tunneling splitting 
$t(\phi)$, and changes with variation of the external induction field. 
This implies that the dipolar coupling between two TLS's depends also on $B$,
and may lead to level crossings. Energy levels of the coupled system 
may cross when their asymmetry is unfavourable to the interaction, and when
the coupling increases sufficiently with magnetic field. This is the
 case if, e.g., 
one TLS has a small tunneling splitting and hence a large
 dipole moment. Imagine 
 two dipole moments,  oriented such that the dipolar interaction would 
flip one of them. When only at higher fields the interaction
is strong enough to achieve this, the first excited state and the ground state
will cross in energy as function of $B$. Thus, a flip of a large total dipole
moment of a coupled ring system is possible.
Both level crossing and strong coupling result in quantum interference effects
on a mesoscopic scale in glasses. At ultra-low temperature one may even expect
a phase transition of coupled TLS's to occur as suggested in \cite{Strehlow}.
We also expect an observable action of the electric flux on the low-temperature
dielectric response of glasses in alternating electrical fields \cite{rogge}.  
In conclusion,  we derived magnetic flux effects in glasses  by considering
low-energy excitations as charged particles moving on closed paths in a double-well
potential. Flux periodic energy levels of these tunneling states 
result in persistent
tunneling currents. Due to strong coupling and level crossing
 magnetic flux effects are strongly enhanced below 100 mK and become 
measurable.

\end{document}